\documentstyle[12pt]{article}
\renewcommand{\theequation}{\arabic{equation}}
\newcommand{\EQ}{\begin{equation}}
\newcommand{\EN}{\end{equation}}

\newcommand{\ket}[1]{\left|#1\right\rangle}      

\newcommand{\bear}{\begin{eqnarray}}
\newcommand{\ear}{\end{eqnarray}}
\newcommand{\bt} { \begin{tabular} }
\newcommand{\et}{ \end{tabular} }
\newcommand{\bc} { \begin{center} }
\newcommand{\ec}{ \end{center} }

\newcommand{\btb} { \begin{table} }
\newcommand{\etb}{ \end{table} }

\begin{document}

\topmargin 0pt
\oddsidemargin 5mm
\newcommand{\NP}[1]{Nucl.\ Phys.\ {\bf #1}}
\newcommand{\PL}[1]{Phys.\ Lett.\ {\bf #1}}
\newcommand{\NC}[1]{Nuovo Cimento {\bf #1}}
\newcommand{\CMP}[1]{Comm.\ Math.\ Phys.\ {\bf #1}}
\newcommand{\PR}[1]{Phys.\ Rev.\ {\bf #1}}
\newcommand{\PRL}[1]{Phys.\ Rev.\ Lett.\ {\bf #1}}
\newcommand{\MPL}[1]{Mod.\ Phys.\ Lett.\ {\bf #1}}
\newcommand{\JETP}[1]{Sov.\ Phys.\ JETP {\bf #1}}
\newcommand{\TMP}[1]{Teor.\ Mat.\ Fiz.\ {\bf #1}}

\renewcommand{\thefootnote}{\fnsymbol{footnote}}

\newpage
\setcounter{page}{0}
\begin{titlepage}
\begin{flushright}
UFSCARF-TH-07-3
\end{flushright}
\vspace{0.5cm}
\begin{center}
{\large The Bethe ansatz approach for factorizable 

centrally extended $\bf{su}(2|2)$ $S$-matrices}\\
\vspace{1cm}
{\large M.J. Martins and C.S. Melo} \\
\vspace{1cm}
{\em Universidade Federal de S\~ao Carlos\\
Departamento de F\'{\i}sica \\
C.P. 676, 13565-905~~S\~ao Carlos(SP), Brasil}\\
\end{center}
\vspace{0.5cm}

\begin{abstract}
We consider the Bethe ansatz solution of integrable models interacting through
factorized $S$-matrices based on the central extention of the
$\bf{su}(2|2)$ symmetry.
The respective $\bf{su}(2|2)$ $R$-matrix
is explicitly related to that of the covering Hubbard model through a 
spectral parameter dependent transformation. This mapping allows us to 
diagonalize  inhomogeneous 
transfer matrices whose statistical weights are given 
in terms of $\bf{su}(2|2)$ $S$-matrices 
by the algebraic Bethe 
ansatz. As a consequence of that we derive the quantization condition on the circle  
for the asymptotic momenta of particles scattering by the $\bf{su}(2|2) \otimes \bf{su}(2|2)$ $S$-matrix.
The result for the quantization rule may be of relevance in the study of the
energy spectrum of the $AdS_5 \times S^{5}$ string sigma model in the thermodynamic limit.
\end{abstract}

\vspace{.15cm}
\centerline{PACS numbers:  05.50+q, 02.30.IK}
\vspace{.1cm}
\centerline{Keywords: Bethe Ansatz, S-matrix }
\vspace{.15cm}
\centerline{March 2007}
\end{titlepage}

\renewcommand{\thefootnote}{\arabic{footnote}}

\section{Introduction}

Nowadays there exists a considerable amount of evidences 
that integrable structures appear in both planar ${\cal{N}}=4$ supersymmetric Yang-Mills
theory \cite{MI,SA} and free IIB superstring theory on 
the $AdS_{5}\times S^{5}$ curved space \cite{RO,OT}. 
Integrability in the 
planar ${\cal{N}}=4$ gauge theory\footnote{For a review on integrable properties
of general four-dimensional gauge theories see \cite{BEL}.} is related 
to the fact that the spectrum of the anomalous dimensions of the conformal
operators can be obtained by diagonalizing long-range integrable spin chains \cite{BE1,BE2}.
For the $AdS_{5}\times S^{5}$ string theory, signatures of integrability have been found on 
the context of semi-classical string states
\cite{TS,KA,REV}.
There is, however, indications that integrability may 
survive when quantum corrections are taken into account
\cite{IA}. There exists also arguments in favour of 
absence of particle production at tree level for the
$AdS_{5} \times S^{5}$ gauge-fixed sigma model \cite{MC} and that the
respective quantum world-sheet energies may arise 
from quantization conditions of the form of
Bethe ansatz equations \cite{BE3,FR,FR1,KZ}.

Assuming that such gauge and string theories are indeed quantum integrable 
it is natural to expect that much of their properties could be
inferred within the factorized $S$-matrix framework \cite{ZA}.
In the ${\cal{N}}=4$ gauge theory the $S$-matrix is supposed to encode the interactions of 
a long-rang spin chain and it is expected to satisfy the 
dynamical form of the Yang-Baxter equation \cite{BE4,BE5}.
From the $AdS_{5}\times S^{5}$ string theory perspective the $S$-matrix 
should describe the scattering amplitudes between the world-sheet
excitations and therefore has to obey the standard Yang-Baxter equation \cite{FRO1,JA}.
The basic form of these two matrix operators have recently been argued to be constrained
by the requirement that they are invariant under the centrally
extended $\bf{su}(2|2)\otimes \bf{su}(2|2)$ superalgebra \cite{BE4,BE5,FRO1}.
The only remaining freedom consists in an abelian phase factor 
that has been argued to be constrained with the help of crossing symmetry
\cite{JA}.

The next step would then be to  determine 
the quantization conditions for
the particle momenta excitations on a ring of size $L$. This requisite is necessary  
to perform non-perturbative
investigations of the energy and momenta eigenstates
associated to the $AdS_5 \times S^{5}$ gauged-fixed sigma model in the infinite volume limit.
The solution of this problem is directly related to the diagonalization of an 
inhomogeneous two-dimensional vertex model of statistical
mechanic whose Boltzmann weights are the matrix elements of the underlying factorized $S$-matrix.
In fact, this task has recently been discussed by Beisert \cite{BE5} in the context of the ${\cal{N}}=4$ gauge theory.
The respective transfer matrix eigenvalues were proposed by 
means of the analytical Bethe ansatz method and apparently also with the help of 
a mapping to the Hubbard model.
We recall that the construction of eigenvectors is unfortunately beyond the scope of the analytical Bethe ansatz.

It seems therefore of interest to tackle the problem of 
diagonalization of transfer matrices based on centrally extended $\bf{su}(2|2)$
$S$-matrices by an independent, first principle approach, such as the algebraic Bethe ansatz method \cite{FA,KO}.
This framework 
can provide us in an 
unambiguous way both the transfer matrix eigenvalues and eigenvectors as well as the respective Bethe ansatz
equation.
To this end, the string theory point of view appears to be the ideal one, 
since the respective $S$-matrix satisfies canonical properties
expected from $(1+1)$-dimensional integrable theories \cite{FRO1}.
The exact knowledge of the eigenvectors, in a near future,
could be of utility in the understanding 
of the Hilbert space structure of the
$AdS_{5}\times S^{5}$ string theory in the thermodynamic limit.

In this paper we are going to consider the latter approach in order to present the 
Bethe ansatz solution of the monodromy problem 
associated to the $\bf{su}(2|2)\otimes \bf{su}(2|2)$ string $S$-matrix.
We have organized this paper as follows. In next section we describe the centrally extended $\bf{su}(2|2)$ $R$-matrix in
a convenient basis. We show that this operator is related to the $R$-matrix of the covering Hubbard model by means of
a unitary transformation depending on the spectral parameters.  In section 3, with the help of this mapping,
we formulate
the diagonalization of
an inhomogeneous transfer matrix based on the $\bf{su}(2|2)$ $R$-matrix weights by the algebraic Bethe
ansatz approach. In section 4 we use this result to derive the quantization rule on a ring of
size $L$ for the particle momenta within the asymptotic Bethe ansatz
framework \cite{SU}. Our conclusions are summarized in section 5 and in Appendix A
we have collected Shastry's $R$-matrix.

\section{The $\bf{su}(2|2)$ $R$-matrix}

In this section we are going to discuss a family of solutions of the Yang-Baxter equation,
\begin{equation}
\label{yb}
R_{23}(\lambda_{1},\lambda_{2}) R_{12}(\lambda_{1},\lambda_{3}) R_{23}(\lambda_{2},\lambda_{3})=
R_{12}(\lambda_{2},\lambda_{3}) R_{23}(\lambda_{1},\lambda_{3}) R_{12}(\lambda_{1},\lambda_{2}),
\end{equation}
where the $R$-matrix $R_{ab}(\lambda,\mu)$ acts on the tensor product of two $Z_2$ graded spaces 
$V_a^{(0)} \oplus V_a^{(1)} $ and 
$V_b^{(0)} \oplus V_b^{(1)} $. As usual, the $\alpha$th element 
of the even $V^{(0)}$ and odd $V^{(1)}$ subspaces
are distinguished by its $Z_2$ parity $\theta_{\alpha}$, 
\begin{equation}
\label{grad}
\theta_{\alpha} =\cases{
\displaystyle  0 \;\; \mbox{for} \;\; \alpha \in V^{(0)}~~(\mathrm{even}) \cr
\displaystyle  1 \;\; \mbox{for} \;\; \alpha \in V^{(1)}~~(\mathrm{odd}) \cr} .
\end{equation}

Here we shall consider the case where the  graded vector space 
is that of a central extension of the $\bf{su}(2|2)$ 
superalgebra \cite{SUP} on its fundamental
four-dimensional representation \cite{BE5,YZ}. In what follows, it will be convenient to
represent the $R$-matrix $R_{12}(\lambda,\mu)$ in terms
of its matrix elements and we shall use the following
representation,
\begin{equation}
\label{rmname}
R_{12}(\lambda,\mu)= \sum_{a,b,c,d=1}^{4} R_{b,a}^{c,d}  e_{ac} \otimes e_{bd},
\end{equation}
where $e_{ab}$ denotes standard  Weyl matrices.

One important 
feature of equation (\ref{yb}) is that it is 
insensitive to the parities  of the elements of the $\bf{su}(2|2)$
superalgebra. In principle, 
from a given $R$-matrix 
satisfying  equation (\ref{yb}) it is possible 
to define two types of $S$-matrices \cite{KU}
\footnote{Strictly speaking it is required that $R_{ab}^{cd}(\lambda,\mu)=0$ for any
$\theta_a +\theta_b + \theta_c +\theta_d=1,3$.  This  condition is satisfied by
the $\bf{su}(2|2)$ $R$-matrix to be discussed in this section. }.
One of them is the standard $S$-matrix,
\EQ
\label{sr}
S_{12}(\lambda,\mu)=P_{12}R_{12}(\lambda,\mu),
\EN
while the other possibility is the so-called graded $S$-matrix,
\EQ
\label{sgr}
S_{12}(\lambda,\mu)=P_{12}^{(g)}R_{12}(\lambda,\mu),
\EN
where the standard and the graded permutators are given by,
\EQ
P_{12}=  \sum_{a,b=1}^{4} e_{ab}
\otimes {e}_{ba},~~~~~ 
P_{12}^{(g)}= \sum_{a,b=1}^{4} (-1)^{\theta_{a} \theta_{b}} {e}_{ab}
\otimes e_{ba} . 
\EN

The $S$-matrices defined 
by Eqs.(\ref{sr},\ref{sgr}) satisfy the canonical form of the Yang-Baxter equation,
\begin{equation}
\label{ybc}
S_{12}(\lambda_{1},\lambda_{2}) S_{13}(\lambda_{1},\lambda_{3}) S_{23}(\lambda_{2},\lambda_{3})=
S_{23}(\lambda_{2},\lambda_{3}) S_{13}(\lambda_{1},\lambda_{3}) S_{12}(\lambda_{1},\lambda_{2}),
\end{equation}
where in the case of the graded $S$-matrix the tensor products 
take into account the graduation of the subspaces.

The framework of $R$-matrices, however, will be the most convenient one for the purposes of this paper.
In fact, it is the $R$-matrix that plays the decisive role in an algebraic formulation of the Bethe ansatz.
It will also help us to make a clear relationship between the 
standard $\bf{su}(2|2)$ $S$-matrix discussed by Arutynov et al \cite{FRO1} and
the $R$-matrix of the Hubbard model \cite{SA1}.
This connection is expected, since the existence of 
such mapping has been predicted by Beisert \cite{BE5} in the context of the ${\cal{N}}=4$ gauge theory.
It should be stressed, however, that in this paper we intend to search 
a connection between $R$-matrices satisfying the 
same canonical Yang-Baxter relations (\ref{yb}).

In order to accomplish this 
task one has to perform certain unitary transformations 
in the $\bf{su}(2|2)$ $S$-matrix $S_{12}(\lambda,\mu)$ given by
Arutynov et al \cite{FRO1}. More specifically, we 
shall use the property that the Yang-Baxter equation (\ref{ybc}) is invariant
by spectral dependent transformations of the following type,
\EQ
\label{tra}
\bar{S}_{12}(\lambda,\mu)=G_{1}(\lambda) G_{2}(\mu) S_{12}(\lambda,\mu) G_{1}^{-1}(\lambda) G_{2}^{-1}(\mu),
\EN
where $G(\lambda)$ is an arbitrary invertible matrix.

It turns out that the suitable  transformation we need 
to make
the above mentioned connection can be decomposed as,  
\EQ
\label{tra1}
G(\lambda) =
\pmatrix{
1  &  0  &  0  & 0 \cr
0  &  t(\lambda)  &  0  & 0 \cr
0  &  0  &  t(\lambda)  & 0 \cr
0  &  0  &  0  & 1 \cr}  \times
\pmatrix{
1  &  0  &  0  & 0 \cr
0  &  0  &  0  & 1 \cr
0  &  0  &  1  & 0 \cr
0  &  1  &  0  & 0 \cr} ,
\EN
where $t(\lambda)$ plays the role of an external field.

By transforming the centrally extended 
$\bf{su}(2|2)$ standard $S$-matrix given by 
Arutynov et al \cite{FRO1} according to Eqs.(\ref{tra},\ref{tra1}) and by taking into account Eq.(\ref{sr}) we find
that the corresponding $R$-matrix  can be written as,
\bear
\label{rm}
\bar{R}_{12}(\lambda,\mu) &=& a_{1}(\lambda,\mu) [ {e}_{11}\otimes {e}_{11}+
{e}_{14}\otimes {e}_{41} + {e}_{41}\otimes {e}_{14} + {e}_{44}\otimes {e}_{44} ]
\nonumber\\
&+& a_{2}(\lambda,\mu) [ -{e}_{11}\otimes {e}_{44}+
{e}_{14}\otimes {e}_{41} + {e}_{41}\otimes {e}_{14} - {e}_{44}\otimes {e}_{11} ]
\nonumber\\
&+& a_{3}(\lambda,\mu) [ {e}_{22}\otimes {e}_{22}+
{e}_{23}\otimes {e}_{32} + {e}_{32}\otimes {e}_{23} + {e}_{33}\otimes {e}_{33} ]
\nonumber\\
&+& a_{4}(\lambda,\mu) [-{e}_{22}\otimes {e}_{33}+
{e}_{23}\otimes {e}_{32} + {e}_{32}\otimes {e}_{23} - {e}_{33}\otimes {e}_{22} ]
\nonumber\\
&+& a_{5}(\lambda,\mu) [ {e}_{21}\otimes {e}_{12}+
{e}_{31}\otimes {e}_{13} + {e}_{24}\otimes {e}_{42} + {e}_{34}\otimes {e}_{43} ]
\nonumber\\
&+& a_{6}(\lambda,\mu) [ {e}_{12}\otimes {e}_{21}+
{e}_{13}\otimes {e}_{31} + {e}_{42}\otimes {e}_{24} + {e}_{43}\otimes {e}_{34} ]
\nonumber\\
&+& a_{7}(\lambda,\mu)\frac{1}{t(\lambda) t(\mu)} [-{e}_{12}\otimes {e}_{43}+
{e}_{13}\otimes {e}_{42} + {e}_{42}\otimes {e}_{13} - {e}_{43}\otimes {e}_{12} ]
\nonumber\\
&+& a_{8}(\lambda,\mu) t(\lambda) t(\mu) [-{e}_{21}\otimes {e}_{34}+
{e}_{31}\otimes {e}_{24} + {e}_{24}\otimes {e}_{31} - {e}_{34}\otimes {e}_{21} ]
\nonumber\\
&+& a_{9}(\lambda,\mu) \frac{t(\mu)}{t(\lambda)}[ {e}_{22}\otimes {e}_{11}+
{e}_{22}\otimes {e}_{44} + {e}_{33}\otimes {e}_{11} + {e}_{33}\otimes {e}_{44} ]
\nonumber\\
&+& a_{10}(\lambda,\mu)\frac{t(\lambda)}{t(\mu)} [ {e}_{11}\otimes {e}_{22}+
{e}_{11}\otimes {e}_{33} + {e}_{44}\otimes {e}_{22} + {e}_{44}\otimes {e}_{33} ]
\ear

The ten distinct weights $a_{i}(\lambda,\mu)$ are obtained directly 
from \cite{FRO1} and they are given by,
\bear
a_{1}(\lambda,\mu)&=&\frac{[x^{-}(\mu)-x^{+}(\lambda)]}{[x^{+}(\mu)-x^{-}(\lambda)]} \frac{\eta(\mu)}{\eta(\lambda)}  
\nonumber \\
a_{2}(\lambda,\mu)&=&\frac{[x^{-}(\lambda)-x^{+}(\lambda)] [x^{-}(\mu)+x^{+}(\lambda)] [x^{-}(\mu)-x^{+}(\mu) ] }
{[x^{-}(\lambda)-x^{+}(\mu)] [x^{-}(\lambda) x^{-}(\mu)-x^{+}(\lambda) x^{+}(\mu) ]} \frac{\eta(\mu)}{\eta(\lambda)}
\nonumber \\
a_{3}(\lambda,\mu)& =& -1
\nonumber \\
a_{4}(\lambda,\mu)&=&\frac{[x^{-}(\lambda)-x^{+}(\lambda)] [x^{-}(\mu)-x^{+}(\mu)] [x^{-}(\lambda)+x^{+}(\mu) ] }
{[x^{-}(\lambda)-x^{+}(\mu) ] [x^{-}(\lambda) x^{-}(\mu)-x^{+}(\lambda) x^{+}(\mu)]}
\nonumber \\
a_{5}(\lambda,\mu)&=&\frac{[x^{-}(\mu)-x^{-}(\lambda)]}{[x^{+}(\mu)-x^{-}(\lambda)]} \eta(\mu)
\nonumber\\
a_{6}(\lambda,\mu)&=&\frac{[x^{+}(\lambda)-x^{+}(\mu)]}{[x^{-}(\lambda)-x^{+}(\mu)]} \frac{1}{\eta(\lambda)}
\nonumber\\
a_{7}(\lambda,\mu)&=&\frac{\sqrt{[x^{-}(\lambda)-x^{+}(\lambda)] [x^{-}(\mu)-x^{+}(\mu)]} [x^{+}(\lambda)-x^{+}(\mu)]}
{[x^{-}(\lambda)-x^{+}(\mu)] [x^{-}(\lambda) x^{-}(\mu)-1] \eta(\lambda)}
\nonumber\\
a_{8}(\lambda,\mu)&=&\frac{\sqrt{[x^{-}(\lambda)-x^{+}(\lambda)] [x^{-}(\mu)-x^{+}(\mu)]} [x^{+}(\lambda)-x^{+}(\mu)]}
{[x^{-}(\lambda)-x^{+}(\mu)] [x^{-}(\lambda) x^{-}(\mu)-1] \eta^{2}(\lambda) \eta(\mu)}
\nonumber\\
a_{9}(\lambda,\mu)&=&\frac{\sqrt{[x^{-}(\lambda)-x^{+}(\lambda)] [x^{-}(\mu)-x^{+}(\mu)]}}
{[x^{+}(\mu)-x^{-}(\lambda)]}
\nonumber\\
a_{10}(\lambda,\mu)&=&\frac{\sqrt{[x^{-}(\lambda)-x^{+}(\lambda)] [x^{-}(\mu)-x^{+}(\mu)]}}
{[x^{+}(\mu)-x^{-}(\lambda)]} \frac{\eta(\mu)}{\eta(\lambda)} ,
\label{weights}
\ear
where $\eta(\lambda)=\sqrt{\frac{x^{+}(\lambda)}{x^{-}(\lambda)}}$.
The functions $x^{\pm}(\lambda)$ depend on both the string sigma model 
coupling constant $g$ and the world-sheet rapidity $\lambda$.
They are constrained to satisfy the following relations \cite{BE4,BE5},
\EQ
\label{vinc}
\label{string}
\frac{x^{+}(\lambda)}{x^{-}(\lambda)}=e^{i \lambda},~~~~
~~ x^{+}(\lambda)+\frac{1}{x^{+}(\lambda)}-x^{-}(\lambda)-\frac{1}{x^{-}(\lambda)}=\frac{i}{g}.
\EN

Before proceeding we note that  
the constant part of the transformation (\ref{tra1}) only relabels 
the Weyl basis.  This means that for $t(\lambda)=1$ the operator 
$\bar{S}_{12}(\lambda,\mu)=P_{12} \bar{R}_{12}(\lambda,\mu)$ is just the
original $\bf{su}(2|2)$ $S$-matrix    
discussed by Arutynov et al \cite{FRO1} written 
in a different basis.  We emphasize that
in this new basis  
the Grassmann parities $\theta_{\alpha}$ 
associated to the $R$-matrix (\ref{rm}-\ref{vinc}) is,
\begin{equation}
\label{par}
\theta_{\alpha}=\cases{
1 \;\;\; \mbox{for} \;\; \alpha = 2,3 \;\; \cr
0 \;\;\; \mbox{for} \;\; \alpha = 1,4 \;\; \cr }.
\end{equation}

The first step to  compare
the $R$-matrix (\ref{rm}-\ref{vinc}) with the graded $R$-matrix of the Hubbard model \cite{SA1,WA} is to rewrite the
restriction (\ref{string}) in terms of Shastry's original coupling constraint. Let us denote this latter
$R$ matrix by 
$R_{12}^{(s)}(\lambda,\mu)$ which for sake of completeness has been summarized in Appendix A. It turns out that
a convenient parameterization for the variables $x^{\pm}(\lambda)$ is \footnote{We recall that this type
of parameterization has first appeared in \cite{SA1,PB}  to simplify the transfer matrix eigenvalues of the covering Hubbard model.}
\EQ
\label{param}
x^{+}(\lambda)=i \frac{a(\lambda)}{b(\lambda)} e^{2 h(\lambda)}, 
~~~~~~x^{-}(\lambda)=-i \frac{b(\lambda)}{a(\lambda)} e^{2 h(\lambda)}
\EN
where $a(\lambda)$, $b(\lambda)$ and $h(\lambda)$ are 
the free-fermion weights and the coupling entering Shastry's
$R$ matrix defined by Eq.(\ref{cons}).

By substituting Eq.(\ref{param}) into the $R$-matrix weights 
(\ref{weights}) and by comparing them with Shastry's weights
given by Eqs.(\ref{shawe}) we see that 
a perfect matching  occurs for a special choice of function $t(\lambda)$, namely
\EQ
\label{te}
t(\lambda)=\sqrt{\eta(\lambda)}=\left(\frac{x^{+}(\lambda)}{x^{-}(\lambda)}\right)^{1/4}
\EN

We see that for this particular value of $t(\lambda)$ the Boltzmann weights of the
$R$-matrix (\ref{rm}-\ref{vinc}) are brought to the most possible symmetrical form. For 
the specific value (\ref{te}) we found 
the following relationship,
\EQ
\label{rs}
\frac{\bar{R}_{12}(\lambda,\mu)}
{\bar{R}(\lambda,\mu)_{1,1}^{1,1}}=
\frac{R_{12}^{(s)}(\lambda,\mu)}
{R^{(s)}(\lambda,\mu)_{1,1}^{1,1}}
\EN

The above discussion reveals us that both the centrally 
extended $\bf{su}(2|2)$ $R$-matrix and Shastry's graded $R$-matrix are indeed special
cases of a family of $R$-matrices $\bar{R}_{12}(\lambda,\mu)$ having 
a spectral dependent free-parameter.
This result brings extra support to the claim by Beisert \cite{BE5} that the 
symmetry underlying the integrability of the Hubbard model
should be that of the $\bf{su}(2|2)$ superalgebra.
Indeed, our result makes it precise the way one can go from the $\bf{su}(2|2)$ 
$R$-matrix to that of the Hubbard model within a common integrable
structure satisfying the canonical Yang-Baxter relation (\ref{yb}).
This comparison will be of great help to formulated the 
corresponding algebraic Bethe ansatz solution in next section.

\section{The algebraic Bethe ansatz}

In this section we shall study the problem of diagonalizing an inhomogeneous row-to-row transfer matrix by an algebraic Bethe ansatz,
 \EQ
\label{tran}
 T(\lambda,\lbrace p_{i} \rbrace )\ket{\Phi}=\Lambda(\lambda,\lbrace p_{i} \rbrace )\ket{\Phi}
 \EN
where $p_1, \dots, p_N$ are the inhomogeneities.

We will  consider the situation in which the Boltzmann weights $\bar{S}_{12}(\lambda,\mu)$ of the transfer matrix
$T(\lambda,\lbrace p_{i} \rbrace )$ take into account the graduation of the degrees of freedom,
\EQ
\tilde{S}_{12}(\lambda,\mu)=P_{12}^{(g)} \bar{R}_{12}(\lambda,\mu)
\EN
where the parities of the graded permutator  are defined in Eq.(\ref{par}).

As usual the transfer matrix $T(\lambda,\lbrace p_{i} \rbrace )$ can be written as the supertrace of an operator denominated monodromy
matrix \cite{KU},
\EQ
\label{trans}
T(\lambda,\lbrace p_{i} \rbrace)= {\mathrm{Str}}_{\cal A}[{\cal{T}}_{\cal A}(\lambda,\lbrace p_{i} \rbrace)]=
\sum_{\alpha =1}^{4} (-1)^{\theta_{\alpha}}
 {\cal{T}}_{\alpha \alpha}(\lambda,\lbrace p_{i} \rbrace)
\EN
where ${\cal{T}}_{\alpha \beta}(\lambda,\lbrace p_{i} \rbrace )$ denotes the matrix elements on the auxiliary space
${\cal A} \equiv {\cal C}^{4}$ of the following ordered product of $S$-matrices,
\EQ
\label{monod}
{\cal{T}_{\cal A}}(\lambda,\lbrace p_{i} \rbrace)=\tilde{S}_{{\cal A}N}(\lambda,p_{N}) \tilde{S}_{{\cal A}N-1}(\lambda,p_{N-1}) \dots
\tilde{S}_{{\cal A}1}(\lambda,p_{1})
\EN

An essential ingredient to establish an algebraic Bethe solution is the existence 
of a reference state $\ket{\omega}$ such that the action of
the monodromy operator (\ref{monod}) in this state gives as a result a triangular matrix.
In our case this state is easily built by the following product of local vectors,
\EQ
\ket{\omega}=\prod_{j=1}^{n} \otimes \ket{\omega}_{j}, ~~~~~ \ket{\omega}_{j}=
\pmatrix{
1 \cr
0 \cr
0 \cr
0 \cr}_{j} ,
\EN
which is an exact eigenstate of $T(\lambda,\lbrace p_{i} \rbrace)$.

In order to construct other eigenvectors other than 
$\ket{\omega}$ we need the help of the quadratic algebra satisfied by the monodromy operator,
namely
\EQ
\label{yba}
\bar{R}_{12}(\lambda,\mu) {\cal{T}}(\lambda,\lbrace p_{i} \rbrace) \stackrel{s}{\otimes}{\cal{T}}(\mu,\lbrace p_{i} \rbrace)
={\cal{T}}(\mu,\lbrace p_{i} \rbrace) \stackrel{s}{\otimes}{\cal{T}}(\lambda,\lbrace p_{i} \rbrace) \bar{R}_{12}(\lambda,\mu) ,
\EN
where the symbol $\stackrel{s}{\otimes}$ stands for 
the supertensor product \cite{KU} whose parities are defined in Eq.(\ref{par}).

From this algebra one can in principle derive suitable commutation 
rules between the monodromy matrix elements acting on the quantum space
$\displaystyle \prod_{j=1}^{n} \otimes {\cal C}^{4}$.
The diagonal monodromy operators define the transfer matrix 
eigenvalue problem while the off-diagonal ones play the role of creation and
annihilation fields over the pseudovaccum $\ket{\omega}$.
A convenient representation of ${\cal T} (\lambda,\lbrace p_{i} \rbrace)$ in terms of these fields turns out to be,
\EQ
\label{abcdf}
{\cal T}(\lambda,\lbrace p_{i} \rbrace) =
\pmatrix{
B(\lambda)       &   \vec{B}(\lambda)   &   F(\lambda)   \cr
\vec{C}(\lambda)  &  \hat{A}(\lambda)   &  \vec{B^{*}}(\lambda)   \cr
C(\lambda)  & \vec{C^{*}}(\lambda)  &  D(\lambda)  \cr}_{4 \times 4},
\EN
where $\vec{B}(\lambda)$
($\vec{B^{*}}(\lambda)$) and
$\vec{C^{*}}(\lambda)$
($\vec{C}(\lambda)$) are two-component row (column) vectors,
$ \hat{A}(\lambda)  $ is a $ 2 \times 2 $ matrix.
The fields $\vec{B}(\lambda)$ , $\vec{B^{*}}(\lambda)$  and  $\vec{C}(\lambda)$,
$\vec{C^{*}}(\lambda)$ are creation and annihilation operators 
over the state $\ket{\omega}$, respectively.

The construction of the eigenvectors in terms of the creation fields will depend much on 
the form of the Boltzmann weights of the $R$-matrix
$\bar{R}_{12}(\lambda,\mu)$.
Considering that the main structure of $\bar{R}_{12}(\lambda,\mu)$ 
resembles that of the Hubbard model one expects that the algebraic Bethe
ansatz solution of the eigenvalue problem (\ref{tran}-\ref{monod}) will 
be similar to that developed for the Hubbard chain \cite{PB}. For arbitrary $t(\lambda)$, however,
not all the weights of $\bar{R}_{12}(\lambda,\mu)$ 
are exactly the same as that of Shastry's graded $R$-matrix.
It turns out that some of them are crucial in the construction of the 
eigenvectors and this means that we need to
implement few adaptations
on the results of \cite{PB} before using them in our situation. 
In what follows, we will describe such modifications in terms of the general 
matrix elements $\bar{R}(\lambda,\mu)_{b,a}^{c,d}$ in order to make the construction of \cite{PB}
more widely applicable. 

The structure of the
eigenvectors is that of multiparticle 
states parameterized by variables $\lambda_{1},\dots,\lambda_{m_1}$ that are fixed
by  Bethe ansatz equations.
Formally, they can be written by the following scalar product \cite{PB},
\EQ
\label{psi}
\ket{\Phi}=\vec{\varphi}_{m_{1}}(\lambda_{1},\dots,\lambda_{m_1}).\vec{{\cal F}}\ket{\omega}
\EN
where the components of the vector $\vec{{\cal F}} \in 
\displaystyle \prod_{j=1}^{m_1} \otimes {\cal C}_{j}^{2}$  shall be fixed by a second
Bethe ansatz.
The vector $\vec{\varphi}_{m_1} (\lambda_{1},\dots,\lambda_{m_1})$ 
carries the dependence on the creation fields and obeys a second order
recursion relation given by,
\bear
\label{vect}
&&\vec{\varphi}_{m_1}(\lambda_1,\dots,\lambda_{m_1})=\vec{B}(\lambda_1) \otimes \vec{\varphi}_{m_{1}-1}(\lambda_2,\dots,\lambda_{m_1})+
\sum_{j=2}^{m_1} \frac{\bar{R}(\lambda_1,\lambda_j)_{4,1}^{2,3} }{ \bar{R}(\lambda_1,\lambda_j)_{4,1}^{4,1} } \prod_{\stackrel{k=2}{k \neq j}}^{m_1}
\frac{\bar{R}(\lambda_k,\lambda_j)_{1,1}^{1,1} }{ \bar{R}(\lambda_k,\lambda_j)_{2,1}^{2,1} } \times
\nonumber \\
&&\vec{\xi} \otimes F(\lambda_1) \vec{\varphi}_{m_{1}-2}(\lambda_2,\dots,\lambda_{j-1},\lambda_{j+1},\dots,\lambda_{m_1}) B(\lambda_j) \prod_{k=2}^{j-1}
\frac{\bar{R}(\lambda_k,\lambda_j)_{2,2}^{2,2} }{ \bar{R}(\lambda_k,\lambda_j)_{1,1}^{1,1} } \check{r}_{k~k+1}(\lambda_k,\lambda_j).
\ear

The auxiliary four-dimensional vector 
$\vec{\xi}$  and the $4 \times 4$ $R$-matrix $\check{r}_{12}(\lambda,\mu)$ are given by,
\begin{equation}
\vec{\xi}=\pmatrix{0 & 1 & -1 & 0 \cr } ~~~~~~~
\check{r}_{12}(\lambda,\mu)=\pmatrix{
1 & 0 & 0 & 0 \cr
0 & \bar{a}(\lambda,\mu) & \bar{b}(\lambda,\mu) & 0 \cr
0 & \bar{b}(\lambda,\mu) & \bar{a}(\lambda,\mu) & 0 \cr
0 & 0 & 0 & 1 \cr}
\end{equation}
where the expressions for the elements 
$\bar{a}(\lambda,\mu)$ and $\bar{b}(\lambda,\mu)$ are,
\bear
&&\bar{a}(\lambda,\mu)=\frac{ \bar{R}(\lambda,\mu)_{3,2}^{2,3} 
\bar{R}(\lambda,\mu)_{4,1}^{4,1}-\bar{R}(\lambda,\mu)_{4,1}^{2,3} \bar{R}(\lambda,\mu)_{3,2}^{4,1}}
{\bar{R}(\lambda,\mu)_{2,2}^{2,2} \bar{R}(\lambda,\mu)_{4,1}^{4,1}}
\nonumber \\
&&\bar{b}(\lambda,\mu)=\frac{ \bar{R}(\lambda,\mu)_{3,2}^{3,2} 
\bar{R}(\lambda,\mu)_{4,1}^{4,1}-\bar{R}(\lambda,\mu)_{4,1}^{3,2} \bar{R}(\lambda,\mu)_{3,2}^{4,1}}
{\bar{R}(\lambda,\mu)_{2,2}^{2,2} \bar{R}(\lambda,\mu)_{4,1}^{4,1}}.
\label{barb}
\ear

From the structure of the $R$-matrix (\ref{rm}) we see that 
the eigenvectors dependence on the free-parameter $t(\lambda)$ is  encoded only
through the weight $\bar{R}(\lambda,\mu)_{4,1}^{2,3}$.
The functions $\bar{a}(\lambda,\mu)$ and $\bar{b}(\lambda,\mu)$ does not carry any 
dependence on this extra parameter since it is canceled
out in the product $\bar{R}(\lambda,\mu)_{4,1}^{2,3} 
\bar{R}(\lambda,\mu)_{3,2}^{4,1}$.
From now on the analysis becomes fairly  parallel 
to that already carried out for the Hubbard model \cite{PB}.
In particular, one expects that the auxiliary $R$-matrix $\check{r}_{12}(\lambda,\mu)$ 
should be that of an isotropic six-vertex model.
This can be seen with the help of the following change of variables,
\EQ
\label{string2}
\tilde{\lambda}= x^{+}(\lambda)+\frac{1}{x^{+}(\lambda)}-\frac{i}{2 g}=
x^{-}(\lambda)+\frac{1}{x^{-}(\lambda)}+\frac{i}{2 g}
\EN

After some simplifications it is possible to rewrite 
$\check{r}_{12}(\tilde{\lambda},\tilde{\mu})$ as
\EQ
\label{rnestedtilde}
\check{r}_{12}(\tilde{\lambda},\tilde{\mu})=\frac{1}{\tilde{\mu}-\tilde{\lambda}+\frac{i}{g}} 
\left[ (\tilde{\mu}-\tilde{\lambda})
\sum_{\alpha \beta=1}^{2} {e}_{\alpha \beta} \otimes {e}_{\beta \alpha}+\frac{i}{g} \sum_{\alpha \beta=1}^{2} {e}_{\alpha \beta}
\otimes {e}_{\alpha \beta} \right]
\EN
which is exactly the $R$-matrix of the rational six-vertex model.

As explained in \cite{PB} the state $\ket{\Phi}$ defined by the expressions (\ref{psi}-\ref{barb}) 
becomes an eigenvector of the transfer
matrix $T(\lambda,\lbrace p_{i} \rbrace)$ under the requirement that the 
vector $\vec{\cal F}$ is an eigenstate of yet another inhomogeneous
transfer matrix whose weights are that of the isotropic six-vertex model (\ref{rnestedtilde}).
The form of the eigenvalues $\Lambda (\lambda,\lbrace p_{i} \rbrace)$ 
depends also on extra variables $\mu_1,\dots,\mu_{m_2}$ that are needed in the
diagonalization of such inhomogeneous six-vertex transfer matrix.
Its final expression in terms of the matrix elements is,
\bear
\label{eigvalues}
\Lambda(\lambda,\lbrace p_{i} \rbrace;\lbrace \lambda_{j},\mu_{l} \rbrace)&=&\prod_{i=1}^{N} \bar{R}(\lambda,p_{i})_{1,1}^{1,1}
\prod_{j=1}^{m_1} \frac{\bar{R}(\lambda_j,\lambda)_{1,1}^{1,1}}{\bar{R}(\lambda_j,\lambda)_{2,1}^{2,1}}
\nonumber \\
&-& \prod_{i=1}^{N} \bar{R}(\lambda_j,\lambda)_{2,1}^{2,1}
\prod_{j=1}^{m_1} -\frac{\bar{R}(\lambda,\lambda_j)_{2,2}^{2,2}}{\bar{R}(\lambda,\lambda_j)_{2,1}^{2,1}}
\left[ \prod_{l=1}^{m_2} \frac{1}{\bar{b}(\mu_l,\lambda)} + \prod_{j=1}^{m_1} \bar{b}(\lambda,\lambda_j)
\prod_{l=1}^{m_2} \frac{1}{\bar{b}(\lambda,\mu_l)} \right]
\nonumber \\
&+&\prod_{i=1}^{N} \bar{R}(\lambda,p_{i})_{4,1}^{4,1}
\prod_{j=1}^{m_1} \frac{\bar{R}(\lambda,\lambda_j)_{4,2}^{4,2}}{\bar{R}(\lambda,\lambda_j)_{4,1}^{4,1}}
\ear
provided that the rapidities $\lbrace \lambda_j \rbrace $ and $\lbrace \mu_j \rbrace $ satisfy the nested Bethe ansatz equations,
\bear
\label{bea}
\prod_{i=1}^{N} \frac{\bar{R}(\lambda_j, p_i)_{1,1}^{1,1}}{\bar{R}(\lambda_j, p_i)_{1,2}^{2,1}} & =&
\prod_{\stackrel{k=1}{k \neq j}}^{m_1}-\frac{\bar{R}(\lambda_j,\lambda_k)_{2,2}^{2,2}}{\bar{R}(\lambda_j,\lambda_k)_{1,2}^{2,1}}
\frac{\bar{R}(\lambda_k,\lambda_j)_{1,2}^{2,1}}{\bar{R}(\lambda_k,\lambda_j)_{1,1}^{1,1}}
\prod_{l=1}^{m_2} \frac{1}{\bar{b}(\mu_l,\lambda_j)}  ~~~~j=1,\dots,m_1,
 \nonumber \\
\prod_{j=1}^{m_1} \bar{b}(\mu_l,\lambda_j) & = & 
\prod_{\stackrel{k=1}{k \neq l}}^{m_2} 
\frac{\bar{b}(\mu_l,\mu_k)}{\bar{b}(\mu_k,\mu_l)}  ~~~~
l=1,\dots, m_2.
\ear

From the above expressions we 
see that both the eigenvalues (\ref{eigvalues}) and the Bethe ansatz equations (\ref{bea})
depend only on Boltzmann weights $\bar{R}(\lambda,\mu)_{b,a}^{c,d}$ that are independent of free-parameter $t(\lambda)$.
This is not surprising since we have shown that this family of 
models are related by the transformation (\ref{tra})
which clearly preserves transfer matrix eigenvalues. The eigenvectors, however, carry a non-trivial dependence on the
parameter $t(\lambda)$  explicitly exhibited in expression (\ref{vect}).

For practical purposes it is relevant to present 
the expressions for the eigenvalues
and Bethe ansatz equations in terms
of the kinematical string variables $x^{\pm}(\lambda)$. Since they do not depend on $t(\lambda)$
the simplifications that we need to perform on Eqs.(\ref{eigvalues},\ref{bea}) follow closely that carried out 
for the Hubbard model \cite{PB}. Ommiting here such 
technicalities we find 
the eigenvalues are given by,
\bear
\label{lambdaexplic}
&&\Lambda(\lambda,\lbrace p_{i} \rbrace;\lbrace \lambda_{j},\mu_{l} \rbrace)=\prod_{i=1}^{N} \left[\frac{x^{-}(p_{i})-x^{+}(\lambda)}
{x^{+}(p_{i})-x^{-}(\lambda)}\right] \frac{\eta(p_{i})}{\eta(\lambda)}
\prod_{j=1}^{m_1} \eta(\lambda) \frac{x^{-}(\lambda)-x^{+}(\lambda_j)}{x^{+}(\lambda)-x^{+}(\lambda_j)}
\nonumber\\
&-&\prod_{i=1}^{N} \frac{x^{+}(\lambda)-x^{+}(p_{i})}{x^{-}(\lambda)-x^{+}(p_{i})} \frac{1}{\eta{(\lambda)}}
\left \{
\prod_{j=1}^{m_1} \eta(\lambda) \left[\frac{x^{-}(\lambda)-x^{+}(\lambda_j)}{x^{+}(\lambda)-x^{+}(\lambda_j)} \right]
\prod_{l=1}^{m_2} \frac{x^{+}(\lambda)+\frac{1}{x^{+}(\lambda)}-\tilde{\mu_l}+\frac{i}{2 g}}
{x^{+}(\lambda)+\frac{1}{x^{+}(\lambda)}-\tilde{\mu_l}-\frac{i}{2 g}} \right.
\nonumber\\
&+&
\left.\prod_{j=1}^{m_1} \eta(\lambda)\left[\frac{x^{+}(\lambda_j)-\frac{1}{x^{+}(\lambda)}}{x^{+}(\lambda_j)-\frac{1}{x^{-}(\lambda)}}\right]
\prod_{l=1}^{m_2} \frac{x^{-}(\lambda)+\frac{1}{x^{-}(\lambda)}-\tilde{\mu_l}-\frac{i}{2 g}}
{x^{-}(\lambda)+\frac{1}{x^{-}(\lambda)}-\tilde{\mu_l}+\frac{i}{2 g}} \right \}
\nonumber\\
&+&
\prod_{i=1}^{N} \left[\frac{1-\frac{1}{x^{-}(\lambda) x^{+}(p_i)}}{1-\frac{1}{x^{-}(\lambda) x^{-}(p_i)}}\right]
\left[\frac{x^{+}(p_i)-x^{+}(\lambda)}{x^{+}(p_i)-x^{-}(\lambda)} \right] \frac{\eta(p_i)}{\eta(\lambda)}
\prod_{j=1}^{m_1} \eta(\lambda) \left [ 
\frac{x^{+}(\lambda_j)-\frac{1}{x^{+}(\lambda)}}{x^{+}(\lambda_j)-\frac{1}{x^{-}(\lambda)}} \right ]
\ear
and the corresponding Bethe ansatz equations become,
\bear
\label{betheexp}
\prod_{i=1}^{N} \left[ \frac{x^{+}(\lambda_j)-x^{-}(p_{i})}{x^{+}(\lambda_j)-x^{+}(p_{i})} \right] \eta(p_{i}) &=&
\prod_{l=1}^{m_2} \frac{x^{+}(\lambda_j)+\frac{1}{x^{+}(\lambda_j)}-\tilde{\mu_l}+\frac{i}{2 g}}
{x^{+}(\lambda_j)+\frac{1}{x^{+}(\lambda_j)}-\tilde{\mu_l}-\frac{i}{2 g}}  ~~~~~ j=1,\dots,m_1,
\nonumber \\
\prod_{j=1}^{m_1} \frac{\tilde{\mu}_{l}-x^{+}(\lambda_j)-\frac{1}{x^{+}(\lambda_j)}+\frac{i}{2 g}}
{\tilde{\mu}_{l}-x^{+}(\lambda_j)-\frac{1}{x^{+}(\lambda_j)}-\frac{i}{2 g}} & =& 
\prod_{\stackrel{k=1}{k \neq l}}^{m_2} 
\frac{\tilde{\mu}_{l}-\tilde{\mu}_{k}+\frac{i}{g}}
{\tilde{\mu}_{l}-\tilde{\mu}_{k}-\frac{i}{g}}   ~~~~~l=1,\dots,m_2.
\ear

We conclude this section with the following comment. Our result 
for the eigenvalues (\ref{lambdaexplic}) is consistent
with that proposed by Beisert \cite{BE5} 
provided that one takes into account the following
change of variables $x^{\pm}(\lambda) \rightarrow x^{\mp}(\lambda)$ and $g \rightarrow -g$. 
Though the  Bethe ansatz equations (\ref{betheexp}) are invariant by 
such transformation the ratio $\frac{x^{+}(\lambda)}{x^{-}(\lambda)}$ defining
the pseudo-momenta is now inverted.
While this ambiguity
can appear in analytical Bethe ansatz analysis it certainly does not 
occur in our algebraic Bethe ansatz framework.

\section{The asymptotic Bethe ansatz}

The purpose of this section is to derive the 
quantization rule for the momenta of the particles interacting through a 
$\bf{su(2|2)} \otimes \bf{su}(2|2)$ 
factorizable $S$-matrix by using the asymptotic Bethe ansatz \cite{SU}. 
This form of scattering has been argued to be the one
relevant for the  world-sheet excitations of the
$Ad S_{5} \times S^{5}$ sigma model \cite{BE4,BE5,FRO1}. Formally, 
the respective scattering matrix can be written as,
\EQ
\label{sads5}
S_{12}^{*}(p_1,p_2)= \hat{S}_{12}(p_1,p_2) \otimes \hat{S}_{12}(p_1,p_2),
\EN
where $p_1, \dots, p_N$ denote the momenta of an interacting $N$-particle system.  

The basic structure of the building block 
$S$-matrix $\hat{S}_{12}(p_1,p_2)$ is constrained by the invariance 
of the solution of the Yang-Baxter equation under the centrally extended $\bf{su}(2|2)$
symmetry.  In principle, one can 
choose  either the standard or the graded $\bf{su}(2|2)$
$S$-matrix discussed in section 2. Our context here, however, is that the 
$S$-matrix $S^{*}_{12}(p_1,p_2)$ plays the role of an 
operator describing the monodromy of the $N$-particle wave function.
When periodic boundary conditions are considered one has to take into account 
the different particle statistics under cyclic permutations.
This compatibility condition leads us to 
single out the graded $\bf{su}(2|2)$ $S$-matrix,
\EQ
\label{ssu2}
\hat{S}_{12}(p_1,p_2)=S_0(p_1,p_2) \left [ P^{(g)}_{12} \bar{R}_{12}(p_1,p_2) \right ].
\EN
where $S_0(p_1,p_2)$ is a scalar factor that 
can not be determined on basis
of the $\bf{su}(2|2)$ invariance.

The possible functional form of $S_0(p_1,p_2)$ has been argued to be restricted by 
unitarity and an extention of the crossing property to $S$-matrices depending
on both momenta $p_1$ and $p_2$ \cite{JA}. This factor was first proposed on the
context of the $AdS^{5} \times S^{5}$ string spectrum \cite{FR} and since then 
has been further investigated by several authors \cite{HL,FRO2,BE6}. The general
structure
of this factor is believed to be given by the expression,
\EQ
[S_0(p_1,p_2)]^2= \frac{x^{+}(p_2)-x^{-}(p_1)}
{x^{-}(p_2)-x^{+}(p_1)}
\frac{1-\frac{1}{x^{+}(p_1) x^{-}(p_2)}}
{1-\frac{1}{x^{-}(p_1) x^{+}(p_2)}} [\sigma(p_1,p_2)]^2
\label{dress}
\EN
where 
$\sigma(p_1,p_2)$ is  the so-called string dressing term. 

This dressing factor can in general be expressed in terms
of a standard phase-shift $\sigma(p_1,p_2)= exp[i \theta(p_1,p_2)] $ where the phase $\theta(p_1,p_2)$
is an anti-symmetric function of the two momenta $p_1$ and $p_2$. For recent closed formula representations
of the gauge independent part of this function see for instance \cite{PHA}. Here we recall that
the property $\theta(p_1,p_2)=
-\theta(p_2,p_1)$ together with the expression for the
$R$-matrix given in section 2 implies that the $S$-matrix 
$S_{12}^{*}(p_1,p_2)$ obeys 
the unitarity condition, namely
\EQ
S_{12}^{*}(p_1,p_2)
S_{21}^{*}(p_2,p_1)=  Id
\EN
where $Id$ is the identity matrix.

We now proceed under the 
assumption of the existence of an abelian  
phase $\theta(p_1,p_2)$  such that
the 
$S$-matrix (\ref{sads5}) indeed encodes  
the interactions of the many-body problem associated to the $Ad S_{5} \times S^{5}$
string sigma model. Under this hypothesis it is possible to find the respective 
particles momenta quantization 
by means of the asymptotic Bethe ansatz framework \cite{SU}.
In this method one assumes that there exists particle coordinate regions $|x_i-x_j| \geq R_c$ where the particles do not interact and that
possible off-mass-shell effects can be neglected.
In such asymptotic regions the wave function is simply the linear combination plane waves with asymptotic momenta $ p_1,\dots,p_N $.
More  precisely, 
the state vector, in an asymptotic region where the particles coordinates are ordered as
$0 \leq x_{{\cal{Q}}_1} < x_{{\cal{Q}}_2} < \dots < x_{{\cal{Q}}_N} \leq L$, has the form  of a generalized
Bethe ansatz \cite{YY,OUT},
\EQ
\label{spsi}
\ket{\Psi}= \int dx_1 \dots dx_N \sum_{{\cal{P}}} A_{\sigma_1, \dots ,\sigma_N} ({\cal{P}}|{\cal{Q}}) e^{\left[ i \sum_{j=1}^{N} p_{{\cal{P}}_j}
x_{{\cal{Q}}_j} \right]} \prod_{k=1}^N \psi_{\sigma_k}^{\dag}(x_k) \ket{0},
\EN
where 
$\psi_{\sigma_k}^{\dag}(x_k)$  creates a particle with internal
quantum number $\sigma_k$ on the vacuum $\ket{0}$. The sum runs over all $N!$ permutations 
$\cal{P}$ of numbers $\{ 1, \cdots,N\}$ that index the particles.

The interaction between the particles 
permit them to cross the various regions and
the  corresponding amplitudes
$A_{\sigma_1, \dots ,\sigma_N} ({\cal{P}}|{\cal{Q}})$  are related to one
another by the $S$-matrix (\ref{sads5}).  For instance,
the amplitudes of two distinct regions $({\cal{P}}|{\cal{Q}})$ and  $(\bar{{\cal{P}}}|\bar{{\cal{Q}}})$ 
differing by the permutation of neighboring $ith$ and
$jth$ particles are connected by
\EQ
\label{perm}
A_{\sigma_1, \dots ,\sigma_i,\sigma_j,\dots, \sigma_N} (\bar{{\cal{P}}}|\bar{{\cal{Q}}})=
S^{*}(p_i,p_j)^{\bar{\sigma}_{i},\bar{\sigma}_{j}}_{\sigma_{i},\sigma_{j}}
A_{\sigma_1, \dots ,\bar{\sigma_i},\bar{\sigma_j},\dots, \sigma_N} ({\cal{P}}|{\cal{Q}}),
\EN
where $S^{*}(p_i,p_j)^{\bar{\sigma}_{i},\bar{\sigma}_{j}}_{\sigma_{i},\sigma_{j}}$ 
are the matrix elements of the $S$-matrix (\ref{sads5}).

The  total energy $E$ and momenta $P$ of the Bethe wave function (\ref{spsi}) are given by
the free particle expressions, namely
\EQ
P= \sum_{k=1}^{N} p_k,~~~~~~~
E=\sum_{k=1}^{N} \varepsilon(p_k),
\EN
where $\varepsilon(p_i)$ is the one-particle dispersion relation.

From Eqs.(\ref{spsi},\ref{perm}) one clearly sees that the role of the scattering theory is to provide the conditions to
match the wave function in adjacent free distinct regions.
Therefore, the information that the internal wave function degrees of freedom satisfy both commuting and anticommuting rules of permutation
should then be encoded in the $S$-matrix (\ref{sads5}).
This feature is guaranteed when one takes as $\hat{S}_{12}(p_1,p_2)$ the graded $\bf{su}(2|2)$ $S$-matrix equation (\ref{ssu2}).
The next step is to quantize the asymptotic momenta $p_k$ by imposing periodic boundary conditions to the wave function
(\ref{spsi},\ref{perm}) on a ring of size $L$.
This is accomplished by the successive use of Eq.(\ref{perm}) in order to relate different regions in the configuration space.
The graded approach assures us strictly periodic boundary conditions in all sectors for both bosonic and fermionic variables.
It turns out that the one-particle momenta $p_k$ are required to satisfy the following condition,
\EQ
e^{-i p_k L}= \frac{\Lambda^{*}(\lambda=p_k,\lbrace p_i \rbrace)}{{S^{*}}_{1,1}^{1,1}(p_k,p_k)}, ~~~k=1,\dots,N
\EN
where $\Lambda^{*}(\lambda,\lbrace p_i \rbrace)$ are the eigenvalues of the transfer matrix operator
$T^{*}(\lambda,\lbrace p_i \rbrace)$ given by
\EQ
T^{(*)}(\lambda,\lbrace p_{i} \rbrace)= {\mathrm{Str}}_{\cal A}[ S_{{\cal A}N}^{(*)}(\lambda,p_{N}) S_{{\cal A}N-1}^{(*)}(\lambda,p_{N-1}) \dots
S_{{\cal A}1}^{(*)}(\lambda,p_{1})]
\EN

There is no need of extra effort to compute the eigenvalues $\Lambda^{*}(\lambda,\lbrace p_i \rbrace)$ because
of the tensor product character of the $S$-matrix 
$S_{12}^{(*)}(p_{1},p_2)$. 
They are simply given in terms of the product of the eigenvalues 
of the transfer matrix diagonalized in
section 2. More specifically we have,
\EQ
\Lambda^{*}(\lambda,\lbrace p_{i} \rbrace) = \prod_{i=1}^{N} \left [ S_{0}(\lambda,p_i) \right ]^2
\Lambda(\lambda,\lbrace p_{i} \rbrace;\lbrace \lambda^{(1)}_{j},\mu^{(1)}_{l} \rbrace)
\Lambda(\lambda,\lbrace p_{i} \rbrace;\lbrace \lambda^{(2)}_{j},\mu^{(2)}_{l} \rbrace)
\EN
where $\lbrace \lambda_{j}^{(\alpha)} \rbrace$ and $\lbrace \mu_{j}^{(\alpha)} \rbrace$ 
$\alpha=1,2$ 
denote the Bethe ansatz roots used to diagonalize
the transfer matrix based on the two 
$\bf{su}(2|2)$ $S$-matrices of the tensor product (\ref{sads5}).

Taking into account the explicit expression
given in Eq.(\ref{lambdaexplic}) as well as the structure of the abelian factor (\ref{dress}) 
one finds that 
the nested Bethe ansatz equations for one-particle momenta $p_k$ are,
\bear
e^{i p_k \left(-L+N-\frac{m_1^{(1)}}{2}- \frac{m_1^{(2)}}{2}  \right)} 
& = &  
e^{i P} 
\prod_{\stackrel{i=1}{i \neq k}}^{N} 
\left[ \frac{x^{-}(p_i)-x^{+}(p_k)}{x^{+}(p_i)-x^{-}(p_k)} \right]
\left[ \frac{1-\frac{1}{x^{+}(p_k) x^{-}(p_i)}}
{1-\frac{1}{x^{-}(p_k) x^{+}(p_i)}} \right ]
[\sigma(p_k,p_i)]^2 
\nonumber \\
&& \times \prod_{\alpha=1}^{2} 
\prod_{j=1}^{m_1^{(\alpha)}} \left[ \frac{x^{+}(\lambda^{(\alpha)}_{j})-x^{-}(p_k)}{x^{+}(\lambda^{(\alpha)}_{j})-x^{+}(p_k)} 
\right]~~~~k=1,\dots,N,
\label{bethe1}
\ear
\bear
e^{i \frac{P}{2}} 
\prod_{i=1}^{N} \left[ \frac{x^{+}(\lambda_j^{(\alpha)})-x^{-}(p_{i})}{x^{+}(\lambda_j^{(\alpha)})-x^{+}(p_{i})} \right] &=&
\prod_{l=1}^{m_2^{(\alpha)}} \frac{x^{+}(\lambda_j^{(\alpha)})+\frac{1}{x^{+}(\lambda_j^{(\alpha)})}-\tilde{\mu_l}^{(\alpha)}+\frac{i}{2 g}}
{x^{+}(\lambda_j^{(\alpha)})+\frac{1}{x^{+}(\lambda_j^{(\alpha)})}-\tilde{\mu_l}^{(\alpha)}-\frac{i}{2 g}}  
~~~~j=1,\dots,m_1^{(\alpha)};\alpha=1,2,
\nonumber \\
\label{bethe2}
\ear
\bear
\prod_{j=1}^{m_1^{(\alpha)}} \frac{\tilde{\mu}_{l}^{(\alpha)}-x^{+}(\lambda_j^{(\alpha)})-\frac{1}{x^{+}(\lambda_j^{(\alpha)})}+\frac{i}{2 g}}
{\tilde{\mu}_{l}^{(\alpha)}-x^{+}(\lambda_j^{(\alpha)})-\frac{1}{x^{+}(\lambda_j^{(\alpha)})}-\frac{i}{2 g}} &=&
\prod_{\stackrel{k=1}{k \neq l}}^{m_2^{(\alpha)}} 
\frac{\tilde{\mu}_{l}^{(\alpha)}-\tilde{\mu}_{k}^{(\alpha)}+\frac{i}{g}}
{\tilde{\mu}_{l}^{(\alpha)}-\tilde{\mu}_{k}^{(\alpha)}-\frac{i}{g}}   ~~~~~l=1,\dots,m_2^{(\alpha)};\alpha=1,2.
\nonumber \\
\label{bethe3}
\ear

Interesting enough, part of the Bethe ansatz equations depend on the total momenta
of a given sector as well as of the number of the Bethe rapidities $N$ and $m_1^{(\alpha)}$.
The latter feature suggests that
$L^{'}=L-N
+\frac{m_1^{(1)}}{2}
+\frac{m_1^{(2)}}{2}$ play the role of an 
effective scale from which densities should probably
be measured.   It turns out that this scale hides the angular momenta 
on $S^{5}$ of the $AdS_5 \times S^{5}$ theory in the light-cone gauge \cite{FRO2}.
This fact can be viewed by bringing the compact Bethe equations (\ref{bethe1}-\ref{bethe3}) close
to the form of those originally proposed by Beisert and Staudacher \cite{BE3}. We start the 
analysis by Eq.(\ref{bethe1}) that carries the explicit dependence on the size of the system.
This equation is translated to that written by Beisert and Staudacher \cite{BE3} by using
the following roots identification,
\EQ
\bullet~x^{\pm}(p_k) = \frac{x^{\pm}_{4,k}}{g},~~ k=1,\dots,K_4  
\EN
\EQ
\bullet~x^{+}(\lambda_j^{(1)}) = \frac{g}{x_{1,j}},~~j=1,\dots,K_1,~~~~~~ 
\bullet~x^{+}(\lambda_{K_1+j}^{(1)})= \frac{x_{3,j}}{g},~~j=1,\dots,K_3 
\EN
\EQ
\bullet~x^{+}(\lambda_j^{(2)}) = \frac{x_{5,j}}{g},~~j=1,\dots,K_5,~~~~~~ 
\bullet~x^{+}(\lambda_{K_5+j}^{(2)})= \frac{g}{x_{7,j}},~~j=1,\dots,K_7 
\EN
where $N \equiv K_4$, $m_1^{(1)} \equiv K_1+K_3$ and $m_1^{(2)} \equiv K_5+K_7$. We recall
that the rapidities $x^{\pm}_{4,k}$ and $x_{i,j}$ for $i=1,3,5,7$ denote the Bethe roots used 
by Beisert and Staudacher \cite{BE3}.

By taking into account the above relabeling of Bethe roots it follows that
Eq.(\ref{bethe1}) is equivalent to,
\bear
e^{-i p_k \left(L-K_4+\frac{K_3-K_1}{2}+\frac{K_5-K_7}{2}  \right)} 
& = &  
e^{i P} 
\prod_{\stackrel{i=1}{i \neq k}}^{K_4} 
\left[ \frac{x^{+}_{4,k}-x^{-}_{4,i}}{x^{-}_{4,k}-x^{+}_{4,i}} \right]
\left[ \frac{1-\frac{g^2}{x^{+}_{4,k} x^{-}_{4,i}}}
{1-\frac{g^2}{x^{-}_{4,k} x^{+}_{4,i}}} \right ]
[\sigma(p_k,p_i)]^2 
\nonumber \\
&& \times 
\prod_{j=1}^{K_3}  
\frac{x^{-}_{4,k}-x_{3,j}}{x^{+}_{4,k}-x_{3,j}} 
\prod_{j=1}^{K_1}  
\frac{1-\frac{g^2}{x^{-}_{4,k} x_{1,j}}}
{1-\frac{g^2}{x^{+}_{4,k} x_{1,j}}} 
\nonumber \\
&& \times 
\prod_{j=1}^{K_5}  
\frac{x^{-}_{4,k}-x_{5,j}}{x^{+}_{4,k}-x_{5,j}} 
\prod_{j=1}^{K_7}  
\frac{1-\frac{g^2}{x^{-}_{4,k} x_{7,j}}}
{1-\frac{g^2}{x^{+}_{4,k} x_{7,j}}} ,~~~~k=1,\dots,K_4
\label{bet1}
\ear

Before proceeding we note that the 
right-hand side of Eq.(\ref{bet1}) reveals us the presence  of the angular momenta
charge $J= \bar{L}-K_4 +\frac{(K_3-K_1)}{2} +\frac{(K_5-K_7)}{2}$ \cite{BE3}. Recall that in 
the notation of \cite{BE3}
the lenght $\bar{L}$ refers to 
effective scale of the system 
conjugated to the momenta variables. Considering the momenta ambiguity 
mentioned in section 3 one has to set here $-J=L$.
This fact is in aggrement with the expected meaning of $J$ as the world sheet thermodynamic scale
of the gauge-fixed  
$AdS_5 \times S^{5}$ model \cite{FRO2}. In order to complete the mapping of the nested Bethe
equations
(\ref{bethe2},\ref{bethe3}) one has to make the following extra identifications, 
\EQ
\bullet~{\tilde{u}}_j^{(1)}= \frac{u_{2,j}}{g},~~ j=1,\dots,K_2 \equiv m_2^{(1)}
~~~~~~~\bullet~{\tilde{u}}_j^{(2)} = \frac{u_{6,j}}{g},~~ j=1,\dots,K_6 \equiv m_2^{(2)}
\EN
as well as the definition $u_{i,j}= x_{i,j}+\frac{g^2}{x_{i,j}}$ for $i=1,3,5,7$.

By using all these identifications, Eqs.(\ref{bethe2}) can be transformed to the following
four type of Bethe equations,
\bear
e^{-i P/2} 
\prod_{i=1}^{K_4}  
\frac{1-\frac{g^2}{x^{-}_{4,i} x_{1,j}}}
{1-\frac{g^2}{x^{+}_{4,i} x_{1,j}}} & =& 
\prod_{l=1}^{K_2} \frac{u_{1,j}-u_{2,l}+\frac{i}{2}}{u_{1,j}-u_{2,l}-\frac{i}{2}},~~~~j=1,\dots,K_1
\nonumber \\
e^{i P/2} 
\prod_{i=1}^{K_4}  
\frac{x^{-}_{4,i}-x_{3,j}}{x^{+}_{4,i}-x_{3,j}} &=&
\prod_{l=1}^{K_2} \frac{u_{3,j}-u_{2,l}+\frac{i}{2}}{u_{3,j}-u_{2,l}-\frac{i}{2}},~~~~j=1,\dots,K_3
\nonumber \\
e^{i P/2} 
\prod_{i=1}^{K_4}  
\frac{x^{-}_{4,i}-x_{5,j}}{x^{+}_{4,i}-x_{5,j}} & =&
\prod_{l=1}^{K_6} \frac{u_{5,j}-u_{6,l}+\frac{i}{2}}{u_{5,j}-u_{6,l}-\frac{i}{2}},~~~~j=1,\dots,K_5
\nonumber \\
e^{-i P/2} 
\prod_{i=1}^{K_4}  
\frac{1-\frac{g^2}{x^{-}_{4,i} x_{7,j}}}
{1-\frac{g^2}{x^{+}_{4,i} x_{7,j}}} & =& 
\prod_{l=1}^{K_6} \frac{u_{7,j}-u_{6,l}+\frac{i}{2}}{u_{7,j}-u_{6,l}-\frac{i}{2}},~~~~j=1,\dots,K_7
\label{bet2}
\ear
while Eqs.(\ref{bethe3}) become,
\bear
\prod_{j=1}^{K_1} \frac{u_{2,l}-u_{1,j}+\frac{i}{2}}
{u_{2,l}-u_{1,j}-\frac{i}{2}}
\prod_{j=1}^{K_3} \frac{u_{2,l}-u_{3,j}+\frac{i}{2}}
{u_{2,l}-u_{3,j}-\frac{i}{2}} & =& 
\prod_{\stackrel{k=1}{k \neq l}}^{K_2} 
\frac{u_{2,l}-u_{2,k}+i}
{u_{2,l}-u_{2,k}-i},~~~~l=1,\dots,K_2
\nonumber \\
\prod_{j=1}^{K_5} \frac{u_{6,l}-u_{5,j}+\frac{i}{2}}
{u_{6,l}-u_{5,j}-\frac{i}{2}}
\prod_{j=1}^{K_7} \frac{u_{6,l}-u_{7,j}+\frac{i}{2}}
{u_{6,l}-u_{7,j}-\frac{i}{2}} & =& 
\prod_{\stackrel{k=1}{k \neq l}}^{K_6} 
\frac{u_{6,l}-u_{6,k}+i}
{u_{6,l}-u_{6,k}-i},~~~~l=1,\dots,K_6
\label{bet3}
\ear

Direct comparison between Eqs.(\ref{bet1},\ref{bet2},\ref{bet3}) for $P=0$ with those presented 
by Beisert and Staudacher in table 5 of reference \cite{BE3} shows us that they are
indeed equivalent in the case of the grading $\eta_1=\eta_2=+1$ choice \footnote{ We also
recall that here $g=\sqrt{\bar{\lambda}}/(4 \pi)$ where $\bar{\lambda}$ is the
't Hooft coupling.}. Here we emphasize
that on the context of 
the $AdS_5 \times S^{5}$
string, however, is 
the charge $J$ that plays 
the role of the thermodynamic scale.  As a consequence of that the thermodynamic limit
$J \rightarrow \infty$ has to be taken without making reference to any particular
sector of the theory. This is of special importance in nested Bethe ansatz systems
since all the Bethe roots levels can in principle contribute to the physical
properties in the infinite volume limit. Recall that this feature has recently
been pointed out to be relevant to unveil the possible origin of the dressing phase
of the ${\cal{N}}=4$ Yang-Mills theory \cite{SAT}.

\section{Conclusion}

In this work we have studied Bethe ansatz properties of integrable models 
associated to centrally extended $\bf{su}(2|2)$ superalgebras. The $\bf{su}(2|2)$
$R$-matrix has been shown to be in the same 
family of Shastry's $R$-matrix \cite{SA1,WA}
with the help
of spectral parameter transformation. This connection made possible 
the solution of the eigenvalue problem associated to the $\bf{su}(2|2)$
transfer matrix within the algebraic Bethe ansatz method.

The motivation to study this type of system arose from its conjectured
relationship with scattering properties of the world-sheet excitations
of the $AdS_5 \times S^{5}$ string sigma model. In this context  we have
been able to derive the Bethe ansatz equations for the respective
particle momenta on the circle. They presented an unusual dependence on
the total momenta and the number of certain rapidities that parameterize
the Hilbert space. We have argued that the latter feature encodes the information
that the effective $AdS^{5} \times S^{5}$ thermodynamic scale is governed
by the angular momenta $J$ charge. This identification allows, in principle,
to take the thermodynamic limit considering the nested Bethe ansatz equations
altogether. 
It remains to be investigate the configurations of the Bethe roots that dominate
the $J \rightarrow \infty$ limit which is crucial for the understanding of
the elementary excitations of the spectrum such as the existence
of possible bound states \cite{MAL,DO}.

\section*{Acknowledgements}
We would like to thank N. Beisert and S. Frolov for useful discussions. 
M.J. Martins would like to thank 
Instituut voor
Theoretische Fysica, Amsterdam, where this work started, for the hospitality. 
This work has been
supported by the Brazilian Research Agencies FAPESP and CNPq.

\addcontentsline{toc}{section}{Appendix A}
\section*{\bf Appendix A: Shastry's $R$-matrix }
\setcounter{equation}{0} \renewcommand{\theequation}{A.\arabic{equation}}

In this appendix we present the explicit expression of 
Shastry's graded $R$ matrix \cite{SA1,WA}. Following the notation
of \cite{PB} this $R$-matrix is given by,
\bear
\label{rmt}
\bar{R}_{12}^{(s)}(\lambda,\mu) &=& \alpha_{2}(\lambda,\mu) [ {e}_{11}\otimes {e}_{11}+{e}_{44}\otimes {e}_{44} ]+
\alpha_{4}(\lambda,\mu) [ {e}_{11}\otimes {e}_{44} + {e}_{44}\otimes {e}_{11} ]
\nonumber\\
&+& \alpha_{1}(\lambda,\mu) [ {e}_{22}\otimes {e}_{22} + {e}_{33}\otimes {e}_{33} ]+
 \alpha_{3}(\lambda,\mu) [ {e}_{22}\otimes {e}_{33} + {e}_{33}\otimes {e}_{22} ]
\nonumber\\
&+& \alpha_{7}(\lambda,\mu) [ {e}_{14}\otimes {e}_{41} + {e}_{41}\otimes {e}_{14} ]-
 \alpha_{6}(\lambda,\mu) [ {e}_{23}\otimes {e}_{32} + {e}_{32}\otimes {e}_{23} ]
\nonumber\\
&-& i \alpha_{8}(\lambda,\mu) [{e}_{21}\otimes {e}_{12}+
{e}_{24}\otimes {e}_{42} + {e}_{31}\otimes {e}_{13} + {e}_{34}\otimes {e}_{43} ]
\nonumber\\
&-& i \alpha_{9}(\lambda,\mu) [{e}_{12}\otimes {e}_{21}+
{e}_{13}\otimes {e}_{31} + {e}_{42}\otimes {e}_{24} + {e}_{43}\otimes {e}_{34} ]
\nonumber\\
&-& i \alpha_{10}(\lambda,\mu) [ {e}_{12}\otimes {e}_{43}-
{e}_{13}\otimes {e}_{42} - {e}_{42}\otimes {e}_{13} + {e}_{43}\otimes {e}_{12} ]
\nonumber\\
&+& i \alpha_{10}(\lambda,\mu) [ {e}_{21}\otimes {e}_{34}-
{e}_{31}\otimes {e}_{24} - {e}_{24}\otimes {e}_{31} + {e}_{34}\otimes {e}_{21} ]
\nonumber\\
&+& \alpha_{5}(\lambda,\mu) [ {e}_{11}\otimes {e}_{22}+
{e}_{11}\otimes {e}_{33} + {e}_{44}\otimes {e}_{22} + {e}_{44}\otimes {e}_{33} ]
\nonumber\\
&+& \alpha_{5}(\lambda,\mu) [ {e}_{22}\otimes {e}_{11}+
{e}_{22}\otimes {e}_{44} + {e}_{33}\otimes {e}_{11} + {e}_{33}\otimes {e}_{44} ]
\ear

The ten non-null Boltzmann weights are,
\bear
\alpha_{1}(\lambda,\mu) & = &
\left\{e^{[h(\mu)-h(\lambda)]}a(\lambda)a(\mu)+e^{-[h(\mu)-h(\lambda)]}b(\lambda)b(\mu) \right \} \alpha_{5}(\lambda,\mu),
\nonumber \\
\alpha_{2}(\lambda,\mu) & = &
\left \{ e^{-[h(\mu)-h(\lambda)]}a(\lambda)a(\mu)+e^{[h(\mu)-h(\lambda)]}b(\lambda)b(\mu) \right \} \alpha_{5}(\lambda,\mu),
\nonumber \\
\alpha_{3}(\lambda,\mu) & =&
\frac{ e^{[h(\mu)+h(\lambda)]}a(\lambda)b(\mu)
+e^{-[h(\mu)+h(\lambda)]}b(\lambda)a(\mu) }
{a(\lambda)b(\lambda)+a(\mu)b(\mu)} \left \{
\frac{\cosh[h(\mu)-h(\lambda)]}{\cosh[h(\mu)+h(\lambda)]} 
\right \} \alpha_{5}(\lambda,\mu),
\nonumber \\
\alpha_{4}(\lambda,\mu) & = &
\frac{e^{-[h(\mu)+h(\lambda)]}a(\lambda)b(\mu)+e^{[h(\mu)+h(\lambda)]}b(\lambda)a(\mu)}
{a(\lambda)b(\lambda)+a(\mu)b(\mu)}
\left \{ \frac{\cosh(h(\mu)-h(\lambda))}{\cosh(h(\mu)+h(\lambda))} 
\right \} \alpha_{5}(\lambda,\mu),
\nonumber \\
\alpha_{6}(\lambda,\mu)& = &\left \{
\frac{ e^{[h(\mu)+h(\lambda)]}a(\lambda)b(\mu)-e^{-[h(\mu)+h(\lambda)]}b(\lambda)a(\mu) }
{a(\lambda)b(\lambda)+a(\mu)b(\mu)} \right \} [b^{2}(\mu)-b^{2}(\lambda)]
\frac{\cosh[h(\mu)-h(\lambda)]}{\cosh[h(\mu)+h(\lambda)]} 
\alpha_{5}(\lambda,\mu),
\nonumber \\
\alpha_{7}(\lambda,\mu) & =& \left \{
\frac{-e^{-[h(\mu)+h(\lambda)]}a(\lambda)b(\mu)+e^{[h(\mu)+h(\lambda)]}b(\lambda)a(\mu)}
{a(\lambda)b(\lambda)+a(\mu)b(\mu)}
\right \} [b^{2}(\mu)-b^{2}(\lambda)]
\frac{\cosh[h(\mu)-h(\lambda)]}{\cosh[h(\mu)+h(\lambda)]} 
\alpha_{5}(\lambda,\mu),
\nonumber \\
\alpha_{8}(\lambda,\mu) & =&
\left \{ e^{[h(\mu)-h(\lambda)]}a(\lambda)b(\mu)-
e^{-[h(\mu)-h(\lambda)]}b(\lambda)a(\mu) \right \} \alpha_{5}(\lambda,\mu),
\nonumber \\
\alpha_{9}(\lambda,\mu) & = &
\left \{ -e^{-[h(\mu)-h(\lambda)]}a(\lambda)b(\mu)+
e^{[h(\mu)-h(\lambda)]}b(\lambda)a(\mu) \right \} 
\alpha_{5}(\lambda,\mu),
\nonumber \\
\alpha_{10}(\lambda,\mu) & =&
\frac{b^{2}(\mu)-b^{2}(\lambda)}{a(\lambda)b(\lambda)+a(\mu)b(\mu)}
\left \{ \frac{\cosh[h(\mu)-h(\lambda)]}{\cosh[h(\mu)+h(\lambda)]} 
\right \} \alpha_{5}(\lambda,\mu),
\label{shawe}
\ear
where the weight $\alpha_{5}(\lambda,\mu)$ is an overall  
normalization. The 
functions $a(\lambda)$, $b(\lambda)$  and the coupling 
$h(\lambda)$ satisfy the constraints,
\begin{equation}
a^{2}(\lambda)+b^{2}(\lambda)=1 ,~~~~~~~
\sinh \left[2 h(\lambda) \right]=\frac{a(\lambda) b(\lambda)}{2 g}
\label{cons}
\EN

{}

\end{document}